\documentclass[aps,pra,twocolumn,showpacs,superscriptaddress,floats]{revtex4-1}
\usepackage{graphicx,bm,amsmath,amssymb,natbib,url,epsfig}

\begin{document}

\title{Quantum noise and mode nonorthogonality\\in nonhermitian
$\mathcal{PT}$-symmetric optical resonators}
\author{Gwangsu Yoo}
\author{H.-S. Sim}
\affiliation{Department of Physics, Korea Advanced Institute of
Science and Technology, Daejeon 305-701, Korea}
\author{Henning Schomerus}
\affiliation{Department of Physics, Lancaster University, Lancaster LA1 4YB, United Kingdom}

\date{\today}

\begin{abstract}
$\mathcal{PT}$-symmetric optical resonators combine absorbing regions with
active, amplifying regions. The latter are the source of radiation
generated via spontaneous and stimulated emission, which embodies quantum
noise and can result in lasing. We calculate the frequency-resolved output
radiation intensity of such systems and relate it to a suitable measure of
excess noise and mode nonorthogonality. The lineshape differs depending on
whether the emission lines are isolated (as for weakly amplifying, almost
hermitian systems) or overlapping (as for the almost degenerate resonances in
the vicinity of exceptional points associated to spontaneous
$\mathcal{PT}$-symmetry breaking). The calculations are carried out in the
scattering input-output formalism, and are illustrated for a quasi
one-dimensional resonator set-up. In our derivations we also allow for the
more general case of a resonator in which the amplifying and absorbing
regions are not related by symmetry.
\end{abstract}



\maketitle

\section{Introduction}

Recent theoretical and experimental advances in optics
\cite{R.El-Ganainy,Z.H.Musslimani2,K.G.Makris,S.Longhi,%
C.E.Ruter,A.Guo,M.C.Zheng,H.Ramezani,A.A.Sukhorukov,Z.Lin,%
H.Schomerus,S.Longhi2,Y.D.Chong} have raised the prospect to
realize nonhermitian systems with a real spectrum but nonorthogonal wave
functions. These efforts are based on the concept of $\mathcal{PT}$ symmetry,
originally formulated as a variant of quantum mechanics in which potentials
can be complex
\cite{C.M.Bender1,M.Znojil,A.Mostafazadeh,A.Mostafazadeh0,C.M.Bender3,F.G.Scholtz}.
In the optical context such potentials can be realized via absorbing and
amplifying regions. In a $\mathcal{PT}$-symmetric situation, a discrete
unitary operation (such as a reflection or inversion) maps the absorbing onto
the amplifying parts, with matching absorption and amplification rates. On
the level of classical optics, absorption and amplification are related by an
antiunitary time-reversal operation. If the absorption and amplification
rates are low enough, the spectrum of the composed system is real, but beyond
a threshold pairs of complex conjugate eigenvalues appear
\cite{C.M.Bender1,M.Znojil,A.Mostafazadeh,A.Mostafazadeh0,C.M.Bender3,F.G.Scholtz,S.Klaiman,%
S.Bittner}.
This transition (known as spontaneous $\mathcal{PT}$-symmetry breaking) can
be exploited to realize a number of exotic optical effects, such as
unidirectional transmission \cite{K.G.Makris,H.Ramezani,Z.Lin},
absorption-enhanced transmission \cite{A.Guo}, power oscillations
\cite{K.G.Makris,C.E.Ruter}, nonlinear switching \cite{A.A.Sukhorukov}, and
coexistence of lasing and perfect absorption \cite{S.Longhi2,Y.D.Chong}. At
the transition point, eigenvalues coalesce, resulting in an \emph{exceptional
point} where the two eigenmodes become degenerate not only in frequency, but
also share the same wave function
\cite{T.Kato,W.D.Heiss,C.Dembowski,M.V.Berry2}. This singular scenario
receives considerable attention also for optical systems without
$\mathcal{PT}$ symmetry
\cite{S.Y.Lee,J.Wiersig,J.-W.Ryu,B.Dietz2,L.Ge,M.Liertzer}.

In this paper, we investigate how the unavoidable consequences of
\emph{leakage}, \emph{instability}, and \emph{quantum noise} affect the
characteristics of realistic $\mathcal{PT}$-symmetric resonators. In
combination, we find that these effects offer a window to directly access the
signatures of nonhermiticity. In particular, they allow to detect mode
nonorthogonality, which discriminates these systems from ordinary hermitian
systems that possess a real spectrum but feature mutually orthogonal
eigenmodes.

In realistic devices, additional losses arise due to \emph{leakage}, as
radiation needs to be coupled out of the system. Even though these losses
break exact $\mathcal{PT}$-symmetry, signatures of the associated peculiar
spectral characteristics are still present in the complex resonance
frequencies of the open system. The coalescence of eigenvalues at the
spontaneous $\mathcal{PT}$-symmetry breaking transition then translates to
situations where two resonance frequencies approach each other very closely
in the complex plane.

In actively amplifying optical systems, the appearance of real
eigenfrequencies indicates an \emph{instability}, the onset of lasing. The
consequences  for $\mathcal{PT}$-symmetric systems have been explored only
since very recently. In these systems, the lasing threshold is either reached
in the limit of the closed system (if the spectrum in this limit is real)
\cite{H.Schomerus}, or at finite leakiness (if $\mathcal{PT}$ symmetry in the
closed system is spontaneously broken, i.e., beyond an exceptional point)
\cite{S.Longhi2,Y.D.Chong}. In both cases, the system is in practice
stabilized by saturation in the amplifying parts, thereby assuring that the
output intensity remains finite. This saturation provides a physical
mechanism that breaks the balance of amplification and absorption required
for $\mathcal{PT}$ symmetry.

An ordinary laser emits coherent radiation with a narrow emission line that
can be well approximated by a Lorentzian. According to general laser theory
\cite{A.E.SiegmanBook,A.L.Schawlow}, the width $\Delta \omega$ of the
Lorentzian arises due to noise, of which a certain amount, \emph{quantum
noise}, is an unavoidable consequence of the quantum nature of microscopic
emission events. Investigations on purely amplifying (not
$\mathcal{PT}$-symmetric) systems established a direct link of nonhermiticity
and an enhanced line broadening (known as excess noise), which are both
captured by a measure of mode nonorthogonality, the Petermann factor $K$
\cite{K.Petermann,A.E.Siegman,P.Goldberg,G.H.C.New,M.Patra,H.Schomerus1}.
At an exceptional point, $K$ diverges because of the coalescence of resonance
wavefunctions \cite{T.Kato,W.D.Heiss,C.Dembowski,M.V.Berry2}.
$\mathcal{PT}$-symmetric systems offer an ideal venue where the consequences
for the radiated intensity in this singular case can be explored. More
generally, one should expect for such systems that the excess noise provides
a probe of the level of nonhermiticity also away from an exceptional point.

The preceding observations capture our principal motivation for this work. It
is the purpose of this paper to formulate a theory of the quantum noise and
radiation of leaky $\mathcal{PT}$-symmetric optical systems in the full range
of situations far below, near, and beyond the reconfiguration of the spectrum
at an exceptional point, up to the point where the lasing threshold is
reached. This requires a quantum optical treatment, which we base on the
scattering input-output formalism
\cite{M.J.Collett,T.Gruner,C.W.J.Beenakker}, as previously applied to purely
amplifying systems \cite{M.Patra,H.Schomerus1}. Taking the
absorbing parts of the system into account, we establish general relations
for the output intensity as for the previously studied case of a
homogeneously amplifying resonator, but find that this involves a nontrivial
combination of aspects from mode nonorthogonality and excess noise. As one
approaches an exceptional point, the partial intensities of the two
near-degenerate resonances still diverge, but the combined amplitude remains
finite, which signals a change in the lineshape from a Lorentzian to a
squared Lorentzian (as also observed at exceptional points in passive
scattering theory \cite{E.Hernandez,N.J.Kylstra,A.I.Magunov}).

We illustrate these general results on quantum noise for a specific quasi
one-dimensional $\mathcal{PT}$-symmetric resonator, which displays the
generic spectral properties of previously studied resonators
\cite{M.Znojil,Y.D.Chong} and offers additional control via a variable
leakage to the exterior. This application extends the investigation in
Ref.~\cite{H.Schomerus}, which used the input-output formalism to study a
${\mathcal PT}$-symmetric resonator with well isolated resonances and did not
address the relation to mode nonorthogonality. Throughout our derivations, we
also present general expressions that apply to systems with amplifying and
absorbing regions, even when these are not related by $\mathcal{PT}$ symmetry
(as recently investigated, e.g., in
Refs.~\cite{L.Ge,M.Liertzer,Y.D.Chong0,J.Andreasen}).

This work is organized as follows. Section \ref{sec:II}  reviews the spectral
features of closed and open $\mathcal{PT}$-symmetric systems, as well as the
signatures of excess quantum noise for conventional, amplifying resonators,
with the discussion based in both cases on the common framework of (first-
and second-quantized) scattering theory. In Section \ref{sec:III} we adapt
the scattering input-output formalism to  resonators with absorbing parts and
derive general expressions for the output radiation intensity. In the central
Section \ref{sec:IV}, we analyze this radiation near resonance and establish
the link to mode nonorthogonality. Section \ref{sec:V} sees our general
results applied to a specific $\mathcal{PT}$-symmetric resonator set-up. We
first study the classical wave problem and determine the resonance
frequencies and exceptional points. We then analyze the output radiation in
the vicinity of these frequencies and verify the sensitivity to the
nonorthogonality of modes, as well as the emergence of a squared Lorentzian
at the exceptional points. Section \ref{sec:conclusion} contains our
conclusions.

\section{\label{sec:II} Background and open issues}

In this section we briefly review the (`first-quantized', classical-wave)
scattering approach to the determination of eigenfrequencies and resonances
in closed and open nonhermitian $\mathcal{PT}$-symmetric systems, as well as
the application of the (`second-quantized', quantum-optical) scattering
input-output formalism  to the problem of excess noise in purely amplifying
systems. Sections~\ref{sec:III} and \ref{sec:IV} then extend the latter to
partially absorbing systems, including those with $\mathcal{PT}$ symmetry.
Throughout, we identify $\omega$ with the energy of photons (effectively
setting $\hbar\equiv 1$).

\subsection{\label{sec:IIa}Scattering quantization and spectral properties of $\mathcal{PT}$-symmetric systems}

While scattering appears most naturally in the context of transport through
open systems, the scattering framework can also serve as a convenient
technique to determine spectral properties, such as the resonance
eigenfrequencies of these systems, or even the bound-state spectrum of closed
systems \cite{E.Doron,E.B.Bogomolny}. In scattering theory the
eigenfrequencies are determined efficiently using a small number of basis
functions, which are the on-shell (fixed energy or frequency) solutions in
various, artificially separated (and therefore open) parts of the system,
which then are coupled together using their scattering matrices. The
quantization condition follows from the consistency requirement of the
matching conditions. In the present work, the use of scattering theory is
further motivated by the fact that it also can serve as a framework for
introducing quantum noise. Indeed, we will find that the steps in the
derivation of the quantization condition (given here) and in the
implementation of quantum noise in composed systems (given in
Sec.~\ref{sec:III}) closely resemble each other. In this first part of
background material, we therefore review the general ideas of spectral
analysis in scattering theory, and also describe the consequences for systems
with $\mathcal{PT}$ symmetry \cite{H.Schomerus,Y.D.Chong} (for other aspects
of $\mathcal{PT}$-symmetric scattering theory see
Refs.~\cite{F.Cannata,M.V.Berry3,H.F.Jones2,H.Schomerus3,H.Hernandez-Coronado,K.Abhinav}).

\subsubsection{Scattering matrix}

We start with the  definition of the relevant scattering matrix. Consider a
classical wave equation, e.g., the Helmholtz equation
\begin{equation}\label{eq:helm}
\Delta \psi({\bf r})+c^{-2}\omega^2n^2({\bf r})\psi({\bf r})=0
\end{equation}
for TM polarized light in a two-dimensional dielectric resonator, where
scattering, absorption, and amplification enter via the dielectric index
$n({\bf r})$ (here $c$ is the speed of light in vacuum). Assuming ${\rm
Re}\,n>0$, we have ${\rm Im}\,n<0$ in amplifying regions, while in absorbing
regions ${\rm Im}\,n>0$. The wave equation is then solved for given
amplitudes ${\bf a}^{\rm in}$ of incident propagating modes, which delivers a
linear relation
\begin{equation}\label{eq:inout1}
{\bf a}^{\rm out}= S(\omega){\bf a}^{\rm in}
\end{equation}
for the amplitudes ${\bf a}^{\rm out}$ of outgoing modes. Here $S(\omega)$ is
the scattering matrix, whose dimensions depend on the number of incoming and
outcoming propagating modes at the given frequency $\omega$.
For a system where modes are coupled in from a left or right entrance, the
scattering amplitudes can be collected into reflection blocks $r$ and $r'$,
as well as transmission blocks $t$ and $t'$ (where the prime discriminates
whether the incident radiation comes from the left or right, respectively),
such that
\begin{equation}\label{eq:sdef}
S=\left(
      \begin{array}{cc}
        r & t' \\
        t & r' \\
      \end{array}
    \right).
\end{equation}

In the special case that amplification and absorption are absent (so that the
refractive index $n$ is real), and if the frequency is real as well, the
scattering matrix is unitary, $S^\dagger(\omega) S(\omega)=\openone$. This
relation embodies the conservation of particle flux. For a complex refractive
index, however, this conservation law is in general violated.

\begin{figure}[t]
\includegraphics[width=\columnwidth]{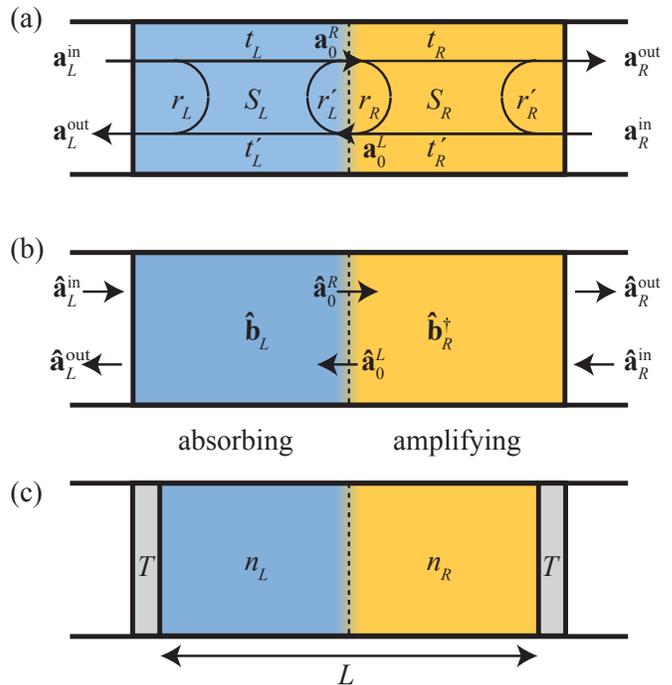}
\caption{\label{fig1}(Color online)
Illustration of the wave-classical and quantum-optical treatment of
$\mathcal{PT}$-symmetric systems, which are composed of an absorbing and an
amplifying region. Panel (a) defines amplitudes used in the scattering
approach to determine the resonance frequencies in the wave-classical limit
(see Sec.~\ref{sec:IIa}). Panel (b) defines the field operators for the
quantum-optical input-output formalism of Sec.~\ref{sec:III}. In panel (c) we
sketch the quasi one-dimensional model resonator studied in Sec.~\ref{sec:V}.
This system is terminated by semitransparent mirrors with transmission
probability $T$. In the interior, two regions of equal length $L/2$ are
equipped with a refractive index $n_L$ (left region) and $n_R$ (right
region), which is constant throughout each respective region. $\mathcal{PT}$
symmetry is realized for $n_R=n_L^*$. We write $n_L=n_0(1+i\alpha)$, where
$\alpha$ controls the degree of nonhermiticity.}
\end{figure}

\subsubsection{Spectral properties of closed systems}

Consider a closed resonator that is split (if only artificially) into two
parts L and R (`left' and `right'), which are described by scattering
matrices identical to the reflection blocks $r_L'(\omega)$ and $r_R(\omega)$,
respectively [see the central region in Fig.~\ref{fig1}(a), ignoring for the
moment any leakage from the system]. At the interface, amplitudes of modes
travelling `to the right' (i.e. from L to R) are collected into a vector
${\bf a}_0^R$, while those for modes travelling into the opposite direction
are collected into a vector ${\bf a}_0^L$. The matching conditions
\begin{equation}
 {\bf a}_0^R=r_L'(\omega){\bf a}_0^L,\quad {\bf a}_0^L=r_R(\omega){\bf a}_0^R
 \label{eq:matching}
\end{equation}
are consistent if
\begin{equation}
{\rm det}\,[r_L'(\omega)r_R(\omega)-\openone]=0.
\label{eq:rlrr}
\end{equation}
For a hermitian system, where the scattering matrices are unitary, this
condition only admits real frequencies. In presence of absorption or
amplification, however, where the scattering matrices are nonunitary, the
eigenfrequencies generally are complex. In both cases, these solutions are
identical to the eigenvalues of the (possibly nonhermitian) underlying
Hamiltonian.

The interest in $\mathcal{PT}$ symmetry for nonhermitian systems arises
because this provides a mechanism where at least part of the spectrum can
still be real. In the scattering formalism, this is embodied by the relation
\cite{H.Schomerus,Y.D.Chong}
\begin{equation}
r_R(\omega)=\mathcal{PT}r_L'(\omega)=[\{r_L'(\omega^*)\}^*]^{-1}
\label{eq:pts0}
\end{equation}
between the scattering matrices, where the labels L and R now refer to the
symmetry-related subsystems. The quantization condition can then be written
as
\begin{equation}
{\rm det}\,[\{r_L'(\omega)\}^*-r_L'(\omega^*)]=0.
\end{equation}
On the real frequency axis, this reduces to the condition \cite{H.Schomerus}
\begin{equation}\label{eq:closedPT}
{\rm det}\,{\rm Im}\,r_L'(\omega)=0,
\end{equation}
which can be generically fulfilled by varying a single real parameter (i.e.,
$\omega$), as it involves the determinant of a manifestly real matrix.
However, the quantization condition can also have complex solutions, which
then occur in complex-conjugated pairs.

The transition between both situations involves pairs of frequencies that
coalesce on the real axis, and then move into the complex plane, where they
remain related by complex conjugation
\cite{C.M.Bender1,M.Znojil,A.Mostafazadeh,A.Mostafazadeh0,C.M.Bender3,F.G.Scholtz,S.Klaiman,%
S.Bittner}. Typically, this transition is driven by increasing the
nonhermiticity (the trend does not need to be strict; sometimes pairs of
resonance frequencies become real when the nonhermiticity is increased).
However, the transition also depends on the coupling of the absorbing and
amplifying regions, and indeed can be induced by reducing this coupling
\cite{H.Schomerus3,O.Bendix,C.T.West}. (In the trivial limit where the two
regions are decoupled, they both possess a fully complex spectrum.)

While ${\cal PT}$ symmetry leaves wavefunctions of real eigenvalues
invariant, it interchanges those of complex-conjugated pairs, which are
therefore not ${\cal PT}$-symmetric when taken each on their own
\cite{C.M.Bender1,C.M.Bender3}. This entails that at the point of degeneracy
the two wavefunctions have to collapse, thereby resulting in an exceptional
point, which is the generic degeneracy scenario in nonhermitian systems
\cite{T.Kato,W.D.Heiss,C.Dembowski,M.V.Berry2}. For the case of the Helmholtz
equation, this behavior can be quantified on the basis of the
bi-orthogonality relation
\begin{equation}\label{biorth}
\int n^2 \psi_1\psi_2 d{\bf r}=0,
\end{equation}
which holds for any two resonance modes $\psi_1$, $\psi_2$, even if the
refractive index is complex. As one approaches the exceptional point,
$\psi_2\to\psi_1\equiv\psi$ is shared between the two degenerate eigenvalues,
and the wave function becomes self-orthogonal,
\begin{equation}\label{selforth}
\int n^2 \psi^2 d{\bf r}=0 \quad\mbox{(exceptional point)}.
\end{equation}

\subsubsection{Spectral properties of open systems}

For a leaky (geometrically open) system, such as  a resonator confined by
semitransparent mirrors, the eigenfrequencies are generally shifted into the
complex plane, corresponding to resonance frequencies of quasibound  states
with decay rate $-2{\rm Im}\,\omega$. These quasibound states fulfill the
wave equation with purely outgoing boundary conditions. For a passive system,
their decay rates are positive. In the presence of amplification
counteracting the leakage, individual resonances can cross from the lower to
the upper half of the complex plane. This signifies an instability, which in
optics corresponds to the laser threshold
\cite{A.E.SiegmanBook,H.Schomerus,S.Longhi2,Y.D.Chong}.

The complex resonance spectrum can again be obtained from scattering theory
\cite{E.Doron}. Including the leakage channels to the outside, the scattering
matrices of the left and right parts assume a natural block structure
\begin{equation}
S_L=\left(
      \begin{array}{cc}
        r_L & t_L' \\
        t_L & r_L' \\
      \end{array}
    \right),\quad
    S_R=\left(
      \begin{array}{cc}
        r_R & t_R' \\
        t_R & r_R' \\
      \end{array}
    \right), \label{eq:slsr}
\end{equation}
which now also contains scattering amplitudes related to reflection and
transmission from and to the exterior regions [i.e., from the left and right
entrances into these subsystems; see Fig.\ \ref{fig1}(a)]. The internal
matching conditions still only involve the reflection blocks $r_L'$ of the
left system and $r_R$ of the right system, and thus remain of the form
\eqref{eq:matching}. Hence, the quantization condition is still given by
Eq.~\eqref{eq:rlrr}. However, for the open system this condition typically
admits only complex solutions, even if the underlying Hamiltonian is
hermitian or ${\cal PT}$-symmetric.

From the perspective of the open system, the significance of
Eq.~\eqref{eq:rlrr} becomes clear when considering the composed scattering
matrix $S=S_L\circ S_R$ of the total system, which is formed according to the
matrix composition rule
\begin{eqnarray}
&&\left( \begin{array}{cc}r_L&t_L'\\t_L&r_L'\end{array} \right)\circ
\left( \begin{array}{cc}r_R&t_R'\\t_R&r_R'\end{array} \right)
\nonumber\\&&=
\left( \begin{array}{cc}r_L+t_L'\frac{1}{1-r_Rr_L'}r_Rt_L&t_L'\frac{1}{1-r_Rr_L'}t_R'
\\ t_R\frac{1}{1-r_L'r_R}t_L&r_R'+t_R\frac{1}{1-r_L'r_R}r_L't_R'\end{array} \right).
\qquad
\label{eq:comprule}
\end{eqnarray}
These expressions contain the resonant denominators  $1-r_L'r_R$, and thus
diverge when condition (\ref{eq:rlrr}) is met.

In the $\mathcal{PT}$-symmetric case, the scattering matrices in Eq.\
(\ref{eq:slsr}) are furthermore related according to \cite{H.Schomerus}
\begin{eqnarray}
&&S_R(\omega)=\mathcal{PT}S_L(\omega)=\sigma_x [S_L^*(\omega^*)]^{-1}\sigma_x
\nonumber \\
&&=\left.\left(
     \begin{array}{cc}
       \frac{1}{(r_L'-t_Lr_L^{-1}t_L')^*} & [r_L^{\prime -1}t_L\frac{1}{t_L'r_L^{'-1}t_L-r_L}]^* \\[.2cm]
       {}[r_L^{-1}t_L'\frac{1}{t_Lr_L^{-1}t_L'-r_L'}]^* & \frac{1}{(r_L-t_L'r_L^{\prime-1}t_L)^*} \\
     \end{array}
   \right)\right|_{\omega^*},\qquad
\label{eq:pts}
\end{eqnarray}
where $\sigma_x$ in the first line is a Pauli matrix, and the second line
follows from block inversion formulas. The resonance condition can then be
written in the form
\begin{equation}\label{eq:ptquant}
{\rm det}\,(r_L'(\omega)-[r_L'(\omega^*)-t_L(\omega^*)r_L^{-1}(\omega^*)t_L'(\omega^*)]^*)=0.
\end{equation}
The quantization condition \eqref{eq:closedPT}  in the closed system is
recovered for $t_L=t_L'=0$.

When the system is almost closed, the eigenfrequencies are only slightly
shifted downwards into the complex plane. The transition of the spectrum from
real to complex is then replaced by a spectral rearrangement in which complex
resonance frequencies approach each other very closely by moving together
roughly parallel to the real axis, and then depart from each other into the
direction of the imaginary axis \cite{A.Guo,S.Longhi2,Y.D.Chong}. When two
complex frequencies meet in an exact degeneracy, one again reaches an
exceptional point, where the wavefunctions of  two resonance modes collapse
and become self-orthogonal, in accordance to Eq.~\eqref{selforth}.

\subsection{\label{sec:IIb} Excess noise for purely amplifying resonators}

A practical method to probe the spectrum of a resonator is to fill it with an
active, amplifying medium and observe the ensuing emitted radiation
intensity. Far above the lasing threshold, one then observes coherent
radiation with a narrow Lorentzian emission line \cite{A.E.SiegmanBook}
\begin{equation}
I(\omega)=\frac{1}{2\pi}I_{\textrm{total}}\frac{\Delta\omega}{(\omega-\omega_0)^2+\Delta\omega^2/4},
\end{equation}
where the line width $\Delta\omega$ is dictated by noise. For the reference
point of an almost closed purely amplifying resonator (the prototypical
\emph{good-cavity laser}), the quantum limit of the line width is given by
the Schawlow-Townes relation \cite{A.L.Schawlow}
\begin{equation}
\Delta\omega =
\Gamma^2/2I_{\textrm{total}},\label{eq:st1}
\end{equation}
where  $I_{\textrm{total}}=\int I(\omega)\,d\omega$, while $\Gamma$ is the
cold-cavity decay rate in the passive resonator [corresponding to a real
refractive index in Eq.~\eqref{eq:helm}]. This noise is due to spontaneous
emission events, which are incoherent and result in phase fluctuations, while
amplitude fluctuations are suppressed by the feedback from the medium
\cite{A.E.SiegmanBook,P.Goldberg}. In the linear regime just below the lasing
threshold, on the other hand, where the emitted radiation is incoherent,
amplitude and phase fluctuations are both present and equal each other, so
that the Schawlow-Townes relation reads
\begin{equation}\Delta\omega =
\Gamma^2/I_{\textrm{total}}.\label{eq:st2}
\end{equation}

Both versions of the Schawlow-Townes relation require absence of absorption
and assume orthogonal resonator modes, so that crosstalk between photons
emitted into different modes can be ignored. In order to identify corrections
due to the violation  of these assumptions, it is advantageous to concentrate
on the linear regime just below threshold, where the emitted intensity can be
calculated conveniently in the scattering input-output formalism
\cite{M.J.Collett,T.Gruner,C.W.J.Beenakker}. Results can be translated to the
lasing regime by the simple rule that there the quantum-limited line width is
still reduced by a factor of two \cite{P.Goldberg}. In the remainder of the
background section we collect results for the previously studied case of
purely amplifying systems
\cite{K.Petermann,A.E.Siegman,P.Goldberg,G.H.C.New,M.Patra,H.Schomerus1},
focussing only on some key points as the details of the derivation are
encompassed by our more general considerations in Secs.~\ref{sec:III} and
\ref{sec:IV}.

The formalism starts with the classical wave equation, e.g., the Helmholtz
equation \eqref{eq:helm}. For a passive system (with a real refractive index)
the scattering matrix $S(\omega)$ is unitary, $SS^\dagger =\openone$, but in
presence of amplification this condition no longer holds. A physical
consequence is the appearance of quantum noise. When amplitudes are promoted
to annihilation operators $\hat {\bf a}^{\rm in, out}$, Eq.\
(\ref{eq:inout1}) with nonunitary $S$ is not compatible with the bosonic
commutation relations for both sets of operators. This problem can be fixed
by introducing auxiliary bosonic operators $\hat{\bf b}$, such that
\begin{equation}\label{eq:inout2}
\hat {\bf a}^{\rm out}= S(\omega)\hat {\bf a}^{\rm in}+Q(\omega)\hat {\bf b}^\dagger.
\end{equation}
The commutation relations for $\hat {\bf a}^{\rm in}$ and $\hat{\bf a}^{\rm
out}$ are now compatible if  $SS^\dagger-QQ^\dagger =\openone$. This
constraint has the status of a fluctuation-dissipation theorem. It admits
solutions when $Q$ is equipped with at least as many columns as rows, but
does not fix $Q$ completely. However, subject to reasonable physical
assumptions, the constraint is sufficient to calculate the output intensity
of the system in terms of its classical wave scattering properties.

For illustration we consider the case of total population inversion ($\langle
\hat{\bf b}^\dagger \hat{\bf b}\rangle=0$) and no incoming radiation. The
frequency-resolved output intensity is then of the form \cite{M.Patra}
\begin{equation}
I(\omega)=\frac{1}{2\pi}
\langle \hat{\bf a}^{{\rm
out} \dagger}\hat{\bf a}^{\rm
out}\rangle=
\frac{1}{2\pi}{\rm tr}\,QQ^\dagger=\frac{1}{2\pi}{\rm tr}\,(SS^\dagger-\openone).
\label{eq:i0}
\end{equation}
The final expression no longer contains the matrix $Q$ and therefore can be
calculated purely based on the classical wave problem. In the vicinity of
resonant emission frequencies, where the scattering matrix diverges [see
Eqs.~\eqref{eq:rlrr} and \eqref{eq:comprule}], the radiated intensity is
large. Linearizing in the deviation from the resonance condition, one
generally finds a Lorentzian line shape
\begin{equation}
I(\omega)=\frac{K}{2\pi}\frac{\Gamma^2}{(\omega-\omega_0)^2+\Delta\omega^2/4},
\label{eq:lorentzian}
\end{equation}
which features an extra factor $K$ so that the Schawlow-Townes relation
Eq.~\eqref{eq:st2} is replaced by
\begin{equation}
\Delta\omega =
K\Gamma^2/I_{\textrm{total}}.
\end{equation}
The factor $K$ is known as the Petermann factor \cite{K.Petermann}, and can
often be related to mode nonorthogonality
\cite{A.E.Siegman,P.Goldberg,G.H.C.New,M.Patra,H.Schomerus1}.
For the Helmholtz equation (\ref{eq:helm}), this factor takes the explicit
form \cite{H.Schomerus1}
\begin{equation}\label{eq:k}
K=\left|\frac{\int |\psi|^2 {\rm Im}(n^2)\,d{\bf r}}{\int \psi^2 {\rm Im}(n^2)\,d{\bf r}}\right|^2,
\end{equation}
which holds also in presence of inhomogeneity in the refractive index and the
gain. For a homogeneously amplifying resonator, this reduces to Siegman's
original expression \cite{A.E.Siegman}
\begin{equation}
K = \frac{|\int |\psi|^2 d{\bf r}|^2}{|\int \psi^2
d{\bf r}|^2}\label{normal_k}.
\end{equation}

The connection of the excess noise to mode nonorthogonality becomes apparent
when one considers the combination (see also Sec.~\ref{sec:IV})
\cite{H.Schomerus1}
\begin{equation}
K\Gamma^2\approx \omega_0^2\left|\frac{\int |\psi|^2 {\rm Im}(n^2)\,d{\bf r}}{\int \psi^2 n^2\,d{\bf r}}\right|^2,
\label{kgamma}
\end{equation}
which appears in the numerator of Eq.~\eqref{eq:lorentzian}. The
self-orthogonality condition \eqref{selforth} implies that this combination
diverges at an exceptional point. In general, the integral in the denominator
of Eq.~\eqref{kgamma} can be interpreted as a suitably weighted overlap
integral of the left eigenfunction $\psi$ and the right eigenfunction
$\psi^*$, which represents the scalar product between these two types of
eigenfunction (the stated simple relation of left and right eigenfunctions
holds because the potential in the Helmholtz equation  is scalar, but breaks
down in the presence of vector potentials \cite{B.Dietz2}). Furthermore,
expressions \eqref{eq:k} and \eqref{normal_k} both imply $K\geq 1$, where
$K=1$ is only obtained in the limit of a passive, closed system, in which the
wave function is real. In more physical terms, on the other hand, expression
(\ref{eq:k}) signifies that the excess noise is generated in the amplifying
parts of the system, since the passive parts (with real refractive index) do
not contribute to the integrals.

Clearly, this duality of interpretations of the Petermann factor in terms of
mode nonorthogonality and excess quantum noise breaks down in presence of
absorbing parts of the system. The mode nonorthogonality is then also induced
by the absorbing regions with a positive imaginary part of the refractive
index, but these on their own do not create any radiation. One of the main
goals of the present work is to identify the role of the Petermann factor in
the presence of such absorbing elements. We focus on $\mathcal{PT}$-symmetric
systems, as their peculiar spectral properties imply that one can also easily
steer close to an exceptional point, and examine how the total output
intensity is regularized despite a divergent Petermann factor. In the
remainder of this paper, we therefore first set up a theory of quantum noise
in simultaneous presence of absorbing and amplifying regions, and then
identify signatures of nonhermiticity, mode-nonorthogonality, and exceptional
points in the output radiation intensity, followed by an illustration for a
specific $\mathcal{PT}$-symmetric system.

\section{\label{sec:III}Input-output formalism in the presence of absorbing regions}

In this section we extend the scattering input-output formalism described in
Sec.~\ref{sec:IIb} to systems which combine amplifying and absorbing regions.

In simultaneous presence of gain by amplification and loss by absorption, the
input-output relations (\ref{eq:inout2}) modify into \cite{C.W.J.Beenakker}
\begin{equation}
\hat {\bf a}^{\rm out}= S(\omega)\hat {\bf a}^{\rm in}+
Q_{(l)}(\omega)\hat {\bf b}_{(l)}+Q_{(g)}(\omega)\hat {\bf b}_{(g)}^\dagger,
\end{equation}
where $(l)$ refers to the absorbing (lossy) regions and $(g)$ refers to the
amplifying regions (with gain). The commutation relations now deliver the
constraint $SS^\dagger+Q_{(l)}^{}Q_{(l)}^\dagger -Q_{(g)}^{}Q_{(g)}^\dagger
=\openone$, which no longer uniquely relates the coupling strengths (encoded
in the combinations $QQ^\dagger$) to the deviation of the scattering matrix
from unitarity.

In order to circumvent this problem, we assume that the absorbing and
amplifying regions are spatially separated. For $\mathcal{PT}$-symmetric
systems, this assumption is rather natural, as the $\mathcal{P}$ symmetry is
usually of a geometric nature, such as a reflection $x\to-x$ about a plane
perpendicular to the $x$-axis. To be specific, we assign absorption to the
left part of the system and amplification to the right part of the system
[see Fig.\ \ref{fig1}(b)]. This strategy, briefly sketched and applied to a
special case in Ref.\ \cite{H.Schomerus}, works whenever a similar separation
into different regions is present due to the physical composition of the
system, even in absence of symmetries. We therefore consider this slightly
more general case, i.e., we first only assume that the absorbing and
amplifying parts can be separated, and only then impose symmetry relations
between them.

The classical wave scattering from  the right and left parts is described by
the scattering matrices $S_L$ and $S_R$, whose natural block structure given
in Eq.\ (\ref{eq:slsr}). The corresponding input-output relations read
\begin{subequations}\begin{eqnarray}
\left( \begin{array}{c} \hat{\bf a}_L^{\rm out} \\ \hat{\bf a}_0^R \end{array} \right)
&=& S_L\left( \begin{array}{c} \hat{\bf a}_L^{\rm in} \\ \hat{\bf a}_0^L \end{array} \right)
+ \left( \begin{array}{c} Q_L \\ Q_L' \end{array} \right)\hat{\bf b}_L,\\
\left( \begin{array}{c} \hat{\bf a}_0^{L} \\ \hat{\bf a}_R^{\rm out} \end{array} \right)
&=& S_R\left( \begin{array}{c} \hat{\bf a}_0^{R} \\ \hat{\bf a}_R^{\rm in} \end{array} \right)
+ \left( \begin{array}{c} Q_R \\ Q_R' \end{array} \right)\hat{\bf b}_R^\dagger,
\end{eqnarray}\label{eq:inout3}\end{subequations}
where the various vectors of operators are defined in Fig.\ \ref{fig1}(b).
Based on their commutation relations, the coupling strengths to the medium,
\begin{subequations}
\begin{eqnarray}
&&\left( \begin{array}{c} Q_L \\ Q_L' \end{array} \right)(Q_L^\dagger,Q_L^{\prime\dagger})
=\openone- S_LS_L^{\dagger},
\\&& \left( \begin{array}{c} Q_R \\ Q_R' \end{array} \right)(Q_R^\dagger,Q_R^{\prime\dagger})
=S_RS_R^{\dagger}-\openone,
\end{eqnarray}\label{eq:qsrelations}\end{subequations}
are now again related to the scattering properties of the purely absorbing or
amplifying parts of the system.
As in the derivation of Eq.~\eqref{eq:i0}, the relations
(\ref{eq:qsrelations}) are sufficient to calculate the output radiation of
the system in terms of the classical scattering properties encoded in $S_L$
and $S_R$.

Starting from the relations (\ref{eq:inout3}), we algebraically eliminate the
auxiliary operators $\hat{\bf a}_0^{L}$ and $\hat{\bf a}_0^{R}$ at the
interface between the two regions,
\begin{subequations}\begin{eqnarray}
&&\hat{\bf a}_0^L=\frac{1}{1-r_Rr_L'}[t_R'\hat{\bf a}_R^{\rm in}+r_Rt_L\hat{\bf a}_L^{\rm in}+Q_R\hat{\bf b}_R^\dagger
+r_RQ_L'\hat{\bf b}_L],
\nonumber\\
{}
\\
&&\hat{\bf a}_0^R=\frac{1}{1-r_L'r_R}[t_L\hat{\bf a}_L^{\rm in}+r_L't_R'\hat{\bf a}_R^{\rm in}+Q_L'\hat{\bf b}_L
+r_L'Q_R\hat{\bf b}_R^\dagger].
\nonumber\\
\end{eqnarray}\end{subequations}
The annihilation operators for outgoing radiation then follow by substituting
these expressions into
\begin{subequations}\begin{eqnarray}
&& \hat{\bf a}_L^{\rm out}=r_L\hat{\bf a}_L^{\rm in}+Q_L\hat{\bf b}_L+t_L'\hat{\bf a}_0^{L},
\\
&& \hat{\bf a}_R^{\rm out}=r_R'\hat{\bf a}_R^{\rm in}+Q_R'\hat{\bf b}_R+t_R\hat{\bf a}_0^{R}.
\end{eqnarray}\end{subequations}

We are interested in the radiation intensity which originates from the
system, i.e., in absence of external incoming radiation, and therefore demand
\begin{equation}
\langle \hat{\bf a}_{nL}^{{\rm in}\dagger}\hat{\bf a}_{nL}^{\rm in} \rangle=
\langle \hat{\bf a}_{nR}^{{\rm in}\dagger}\hat{\bf a}_{nR}^{\rm in} \rangle=0
\end{equation}
for all incoming modes $n$ (expectation values of all crossterms between
different incoming modes also vanish). Within the medium, we assume
mode-independent expectation values
\begin{equation}
\langle \hat{\bf b}_{nL}^{\dagger}\hat{\bf b}_{nL} \rangle=f_L,\quad
\langle \hat{\bf b}_{nR}\hat{\bf b}_{nR}^{\dagger} \rangle=f_R,
\end{equation}
which can be associated to the excited-state occupations $g_L=f_L/(1+2f_L)$
in the absorbing and $g_R=f_R/(2f_R-1)$ in the population-inverted amplifying
parts of the system, respectively.

Under these conditions, the frequency-dependent output intensity of the
composed system, resolved depending on whether it eventually emerges from the
absorbing or the amplifying part of the system, takes the form
\begin{eqnarray}
I_L(\omega)&=&\frac{1}{2\pi}\langle\hat{\bf a}_L^{{\rm out}\dagger}\cdot
\hat{\bf a}_L^{\rm out}\rangle\nonumber
\\
&=&\frac{f_R}{2\pi}{\rm tr}\,\left[t_L'\frac{1}{1-r_Rr_L'}Q_R\right]^\dagger\left[t_L'\frac{1}{1-r_Rr_L'}Q_R
\right]
\nonumber
\\
&+&\frac{f_L}{2\pi}{\rm tr}\,
\left[
Q_L+t_L'\frac{1}{1-r_Rr_L'}r_RQ_L'
\right]^\dagger
\nonumber
\\
&&\quad\times\left[ Q_L+t_L'\frac{1}{1-r_Rr_L'}r_RQ_L' \right],
\label{eq:il0}
\end{eqnarray}
\begin{eqnarray}
I_R(\omega)&=&\frac{1}{2\pi}\langle\hat{\bf a}_R^{{\rm out}\dagger}\cdot
\hat{\bf a}_R^{\rm out}\rangle\nonumber
\\
&=&\frac{f_L}{2\pi}{\rm tr}\,\left[t_R\frac{1}{1-r_L'r_R}Q_L'\right]^\dagger\left
[t_R\frac{1}{1-r_L'r_R}Q_L'\right]
\nonumber
\\
&+&\frac{f_R}{2\pi}{\rm tr}\,
\left[
Q_R'+t_R\frac{1}{1-r_L'r_R}r_L'Q_R
\right]^\dagger
\nonumber
\\
&&\quad\times\left[Q_R'+t_R\frac{1}{1-r_L'r_R}r_L'Q_R\right],
\label{eq:ir0}
\end{eqnarray}
where all combinations of $Q$ matrices are completely determined via the
relations (\ref{eq:qsrelations}). These expressions imply the appearance of
narrow emission lines for frequencies close to resonance, which occurs
whenever the denominator $1-r_L'r_R$ vanishes, in accordance to the
classical-wave quantization condition (\ref{eq:rlrr}).

For a quasi one-dimensional resonator, the transmission and reflection
matrices reduce to complex numbers. Equations (\ref{eq:il0}) and
(\ref{eq:ir0}) then yield the following more compact expressions,
\begin{eqnarray}
I_L &=&  \frac{1}{2\pi}\frac{|t_L'|^2(|r_R|^2+|t_R'|^2-1)}{|1-r_Rr_L'|^2},\label{I_L}\\
I_R &=&  \frac{1}{2\pi}\left[\frac{|t_Rr_L'|^2(|r_R|^2+|t_R'|^2-1)}{|1-r_Rr_L'|^2}+ |t_R|^2+|r_R'|^2-1
\right.\nonumber\\
&&\left.+ 2\,{\rm Re}\,\frac{ t_R r_L'(r_Rt_R^* + t_R'r_R'^*) }{ 1-r_L'r_R }\right],\label{I_R}
\end{eqnarray}
where we now made explicit use of Eq.\ (\ref{eq:qsrelations}) and assumed
idealized conditions with $f_L=0$ (ground state population in the absorbing
parts) and $f_R=1$ (total population inversion in the amplifying parts).  For
$\mathcal{PT}$-symmetric resonators the amplitudes are again related by Eq.\
(\ref{eq:pts}), and the resonance condition is given by Eq.\
(\ref{eq:ptquant}).

\section{\label{sec:IV}Near-resonant radiation intensity and the role of the Petermann factor}

Starting from Eqs.~\eqref{I_L} and  \eqref{I_R} for the emitted intensity
from a quasi one-dimensional resonator, we now evaluate the total intensity
$I=I_L+I_R$ close to resonance, $\omega\approx\omega_0\equiv{\rm
Re}\,\omega_m$, where the complex resonance frequency $\omega_m$ fulfills
$r_L'(\omega_m)r_R(\omega_m)=1$ according to Eq.~\eqref{eq:rlrr}. Keeping
only resonant terms we have
\begin{eqnarray}
I(\omega)=\frac{1}{2\pi}\frac{|t_R|^2(|t_L'|^2+|t_Rr_L'|^2)}{|1-r_L'r_R|^2}\frac{|r_R|^2+|t_R|^{2}-1}{|t_R|^{2}},
\label{eq:ires}
\end{eqnarray}
where we grouped the terms for later convenience. Our goal is to relate this
expression to properties of the resonance wave function $\psi_m$, which
fulfills the one-dimensional Helmholtz equation
\begin{equation}\label{helm1d}
\psi_m''+k_m^2n^2(x)\psi_m=0
\end{equation}
(with $k_m=\omega_m/c$) subject to purely outgoing boundary conditions. The
derivation  proceeds by a number of technical steps which lead to a compact
final result, Eq.~\eqref{finalresult} below.

\subsection{Near-resonant radiation intensity}

For convenience, we normalize the resonance wave function such that in the
free space ($n=1$) to the right of the resonator
$\psi_m(x)=k_m^{-1/2}\exp(ik_m x)$. Furthermore, we also insert a zero-width
layer with $n=1$ at the interface between the right and the left region,
where the wave function then takes the form
\begin{equation}
\psi_m=t_R^{-1}\exp(ik_mx)k_m^{-1/2}+r_Rt_R^{-1}\exp(-ik_mx)k_m^{-1/2}.
\end{equation}
Here we expressed the amplitudes by the elements of the scattering matrix
$S_R$, Eq.~\eqref{eq:slsr}. Furthermore, in terms of elements of the
scattering matrix $S_L$, the outgoing component in the free space to the left
of the resonator is given by $\psi_m(x)=\alpha k_m^{-1/2}\exp(-ik_mx)$, where
$\alpha=t_L'/t_Rr_L'=r_Rt_L'/t_R$ (the latter equality follows from the
resonance quantization condition).

We now evaluate a number of integrals using Eq.~\eqref{helm1d}, integration
by parts, the condition ${\rm Re}\,\omega_m\gg |{\rm Im}\,\omega_m|$, and
boundary terms matching the stipulated outgoing wave amplitudes. Below,
${\cal L}$ denotes the left part of the resonator, and ${\cal R}$ denotes the
right part of the resonator. We then have
\begin{eqnarray}
&&-\int_{{\cal R}} {\rm Im}\,(k_m^2n^2)|\psi_m|^2
\nonumber\\
&&=
 \int_{{\cal R}} {\rm Im}\,(\psi_m^*\psi_m'')
=
\left.{\rm Im}\,(\psi_m^*\psi_m')\right|_{\partial {\cal R}}
\nonumber\\
&&=\frac{|r_R|^2+|t_R|^{2}-1}{|t_R|^{2}}.
\label{relation1}
\end{eqnarray}

Analogously, upon extending the integral over the whole resonator
\begin{equation}
-\int_{{\cal R}+{\cal L}} {\rm Im}\,k_m^2n^2 |\psi_m|^2
=\frac{|t_L'|^2+|t_Rr_L'|^2}{|t_Rr_L'|^2}.
\label{relation2}
\end{equation}

Finally, we employ a similar integral to investigate the behavior of the wave
function close to resonance, $\omega=kc\approx \omega_0$. Due to the
detuning, the wave function then also possesses incoming components, but as
long as the detuning is small the outgoing components remain approximately
unchanged. Therefore, we can assume that to the left of the resonator,
$\psi(x)\approx  k_m^{-1/2}[\alpha\exp(-ik_mx)+\beta\exp(ik_mx)]$ with
$\alpha$ as given before, while to the right $\psi(x)\approx
k_m^{-1/2}[\exp(ik_mx)+\gamma\exp(-ik_mx)]$.

The incoming components of the wave function can now be extracted via the
following sequence of steps (which starts by linearizing $k^2$ around
resonance):
\begin{eqnarray}
&&2k_m(k-k_m)\int_{{\cal R}+{\cal L}} n^2 \psi_m^2
\nonumber\\
&\approx&\int_{{\cal R}+{\cal L}} (k^2-k_m^2)n^2 \psi_m^2
\nonumber\\
&\approx&
\int_{{\cal R}+{\cal L}} (k^2-k_m^2)n^2 \psi_m\psi
\nonumber\\
&=&
\int_{{\cal R}+{\cal L}} (-\psi''\psi_m+\psi_m''\psi)
\nonumber\\
&=&
\left.[-\psi'\psi_m+\psi_m'\psi]\right|_{\partial ({\cal R}+{\cal L})}
\nonumber\\
&=&
2i(\alpha\beta+\gamma).
\label{relation3}
\end{eqnarray}
Furthermore, in terms of the scattering matrix of the whole system we have
$r\beta +t'\gamma=\alpha$ and $t\beta +r'\gamma=1$, which determines the
coefficients $\beta$ and $\gamma$. We can then employ the scattering matrix
composition rules \eqref{eq:comprule} to express these coefficients in terms
of the elements of $S_L$ and $S_R$. Close to resonance, this reduces to
\begin{eqnarray}
(\alpha\beta+\gamma)\to -1/r'\approx -(1-r_L'r_R)/(r_L't_R^2).
\end{eqnarray}

Based on expressions \eqref{relation1}, \eqref{relation2}, and
\eqref{relation3}, the near-resonant radiation intensity \eqref{eq:ires} can
now be rewritten as
\begin{eqnarray}
I(\omega)&=&\frac{1}{2\pi}
\frac{\int_{{\cal R}} {\rm Im}\,\omega_m^2n^2 |\psi_m|^2
 \int_{{\cal R}+{\cal L}} {\rm Im}\,\omega_m^2n^2|\psi_m|^2}{|\int_{{\cal R}+{\cal L}} \omega_m^2 n^2 \psi_m^2|^2}
\nonumber \\
&&{}\times
\frac{\omega_0^2}{(\omega-\omega_0)^2+\Delta \omega^2/4}.
\label{finalresult}
\end{eqnarray}
This is the central general result in this work. It applies to systems with
amplifying and absorbing regions, including ${\cal PT}$-symmetric and purely
amplifying systems, as is discussed in detail in the next subsection. The
last factor is a Lorentzian of width $\Delta \omega=|2\,{\rm Im}\,\omega_m|$,
centered at $\omega_0=\,{\rm Re}\,\omega_m$. The remaining combination of
integrals resembles the expression \eqref{kgamma}  for a purely amplifying
system. However, one of the integrals in the numerator extends over the whole
system (where hermiticity is broken), while the other is restricted to the
(amplifying) right part of the resonator. Therefore, this expression presents
a mixture of the dual interpretations of the conventional Petermann factor as
a measure of mode nonorthogonality and excess noise. Furthermore, as in
Eq.~\eqref{kgamma} the denominator involves the appropriate overlap integral,
which diverges at an exceptional point as a consequence of the
self-orthogonality relation \eqref{selforth}.

\subsection{\label{sec:IVb}Purely amplifying versus ${\cal PT}$-symmetric systems}

In order to get a hold of the general features encoded in the near-resonant
intensity \eqref{finalresult}, we first describe how one recovers from this
expression the Petermann factor \eqref{eq:k} for a purely amplifying
resonator. In such a system, the breaking of hermiticity via the gain
(encoded in ${\rm Im }\,n<0$) is constrained because this systematically
shifts the resonances upwards in the complex plane. Therefore, the system is
driven to the lasing threshold, which is reached when a resonance approaches
the real axis. For highly excited modes (with large $\omega_0$), this happens
very quickly, for $|{\rm Im}\,n|\sim (\Gamma/\omega_0){\rm Re}\,n\ll {\rm
Re}\,n$, where $\Gamma$ is the cold-cavity rate. One then can use
perturbation theory to relate the required gain to $\Gamma$ via \cite{note1}
\begin{equation}\label{eq:gamma}
\frac{\Gamma}{\omega_0}\approx \left|\frac{\int {\rm Im}\,(\omega_m^2n^2)\psi_m^2}{\int \omega_m^2 n^2 \psi_m^2}\right|
\qquad\mbox{\parbox{3.5cm}{(purely amplifying resonator at threshold).}}
\end{equation}
Here, the refractive index is no longer arbitrary but has to be chosen such
that the resonator is at threshold. With the help of this relation,
Eq.~\eqref{finalresult} turns into Eq.~\eqref{eq:lorentzian}, recovering the
Petermann factor $K$ as defined in \eqref{eq:k}.

For systems that are not purely amplifying, on the other hand, the breaking
of hermiticity does not cause a systematic shift of resonances, as the
absorbing regions counteract the effect of the amplifying regions. In
comparison to the Lorentzian \eqref{eq:lorentzian}, one could then  define a
generalized Petermann factor as
\begin{equation}
K=\frac{\int_{{\cal R}} {\rm Im}\,(\omega_m^2n^2) |\psi_m|^2
\int_{{\cal R}+{\cal L}} {\rm Im}\,(\omega_m^2n^2)|\psi_m|^2}
{|\int_{{\cal R}+{\cal L}}\omega_m^2 n^2 \psi_m^2|^2}\frac{\omega_0^2}{\Gamma^2}.
\label{eq:general Petermann factor}
\end{equation}
The same perturbative treatment that leads to Eq.~\eqref{eq:gamma} entails
more generally that the cold-cavity decay rate $\Gamma$ is related to the
(measurable) line width $\Delta \omega=-2{\rm Im}\,\omega$ according to
\begin{equation}\label{eq:gamma2}
\frac{\Gamma}{\omega_0}=\frac{\Delta \omega}{\omega_0}
+\left|\frac{\int {\rm Im}\,(\omega_m^2n^2)\psi_m^2}{\int \omega_m^2 n^2 \psi_m^2}\right|
\quad\mbox{\parbox{3.6cm}{(absorbing \& amplifying resonator, not necessarily at threshold).}}
\end{equation}
[Equation \eqref{eq:gamma} follows by demanding $\Delta\omega\ll \Gamma$
close to threshold.]

For the specific case of a $\mathcal{PT}$-symmetric resonator, the validity
of this perturbative treatment is limited to the regime before the
exceptional point, as such a point induces degeneracy, and the
self-orthogonality property \eqref{selforth} implies that the denominator in
Eq.~\eqref{eq:gamma2} diverges. Before one reaches the exceptional point,
however, the individual wave functions are $\mathcal{PT}$-symmetric, so that
the integral in the numerator of Eq.\ \eqref{eq:gamma2} vanishes. Therefore,
one can safely approximate
\begin{equation}\label{eq:gamma3}
\Gamma=\Delta \omega\qquad\mbox{\parbox{4.5cm}{($\mathcal{PT}$-symmetric resonators),}}
\end{equation}
which in practice should hold up to very close to the exceptional point.
Beyond the exceptional point, it should initially be reasonable to
approximate $\Gamma$  by the average width the two involved, overlapping
resonances. Typically (and as we will confirm below), the regime far beyond
the exceptional point is physically inaccessible as one quickly reaches the
laser threshold \cite{Y.D.Chong}. Based on these observations,
Eq.~\eqref{finalresult} can be directly applied to specific
$\mathcal{PT}$-symmetric systems, which we illustrate in the next section for
the example of a quasi one-dimensional resonator setup.

\section{\label{sec:V} Application to a $\mathcal{PT}$-symmetric resonator}

\subsection{Model system set-up}

We now consider a specific $\mathcal{PT}$-symmetric resonator [depicted in
Fig.~\ref{fig1}(c)], which is made of two regions of equal length $L/2$
(i.e., the total resonator length is $L$).  In the left (absorbing) part of
the system, the complex refractive index $n_L= n_0+i n_I$ ($n_I>0$), while
$n_R=n_L^*=n_0-in_I$ in the right (amplifying) part of the system. The
resonator is terminated by two identical semitransparent mirrors of
transmission probability $T$. This resonator can be interpreted as an open
version of the system studied in Ref.\ \cite{M.Znojil}. It is also similar to
the resonator studied in Ref.\ \cite{H.Schomerus}, but features some
backscattering at the interface between the regions due to the step in ${\rm
Im}\,n$. Furthermore, apart from the additional mirrors, it is like the
resonator studied in Ref.\ \cite{Y.D.Chong}. These slight modifications are
motivated by specific requirements for our investigation. The system must be
open to study the output radiation, the backscattering facilitates the
appearance of exceptional points, and the mirrors allow to study the limit
$T\to 0$ of an almost closed resonator. Furthermore, the degree of
nonhermiticity can be controlled by changing $n_I$.

\subsection{\label{sec:Va}Resonance frequencies and exceptional points}

In order to obtain the resonance frequencies we apply the scattering
quantization formalism of Sec.\ \ref{sec:IIa}. We compose the scattering
matrices $S_L$ and $S_R$ from the scattering matrix
\begin{equation}
S_T=\left( \begin{array}{cc} -\sqrt{1-T} & -i\sqrt{T} \\ -i\sqrt{T} & -\sqrt{1-T}\end{array} \right)
\end{equation}
of a semitransparent mirror with transmission probability $T$, the scattering
matrix
\begin{equation}
S_{n_1,n_2}=\frac{1}{n_1+n_2}\left( \begin{array}{cc}n_1-n_2 & 2\sqrt{n_1n_2} \\ 2\sqrt{n_1n_2} &n_2-n_1\end{array} \right)
\end{equation}
for a refractive index step from $n_1$ to $n_2$,  and the scattering matrix
\begin{equation}
S_{n}=\left( \begin{array}{cc}0 & \exp(i\omega nL/2c) \\ \exp(i\omega nL/2c) &0\end{array} \right)
\end{equation}
for ballistic propagation through a segment of refractive index $n$ and
length $L/2$. Then \cite{note2}
\begin{subequations}
\begin{eqnarray}
S_L&=&S_T\circ S_{1,n_L}\circ S_{n_L}\circ S_{n_L,1},\label{eq:S_L}
\\
S_R&=&S_{1,n_R}\circ S_{n_R}\circ S_{n_R,1}\circ S_T,\label{eq:S_R}
\end{eqnarray}
\end{subequations}
with matrix composition rule \eqref{eq:comprule}.

Following this prescription, we obtain the elements of $S_L$ in the form
\begin{subequations}\label{eq:sl}
\begin{eqnarray}
&&r_L=-
\frac{(n_L+1)A_L^{+}+(n_L-1)A_L^{-}X_L^2}
{(n_L+1)A_L^{-}+(n_L-1)A_L^{+}X_L^2},
\label{r_L}\\
&&t_L'=t_L=-
\frac{2in_L\sqrt{T}X_L}
{(n_L+1)A_L^{-}+(n_L-1)A_L^{+}X_L^2},
\\
&&r_L'=-
  \frac{(n_L-1)A_L^{-}+
  (n_L+1)A_L^{+}X_L^2
  }
{(n_L+1)A_L^{-}+(n_L-1)A_L^{+}X_L^2},
\end{eqnarray}\end{subequations}
where we introduced
\begin{equation}
A_L^{\sigma}=
\frac{1}{2}[(1+\sigma n_L)\sqrt{1-T}+ 1-\sigma n_L],
\quad X_L=e^{i\omega n_L L/2c}.
\end{equation}
Analogously,
\begin{subequations}\label{eq:sr}
\begin{eqnarray}
&&r_R=-
  \frac{(n_R-1)A_R^{-}+
  (n_R+1)A_R^{+}X_R^2
  }
{(n_R+1)A_R^{-}+(n_R-1)A_R^{+}X_R^2},
\\
&&t_R'=t_R=-
\frac{2in_R\sqrt{T}X_R}
{(n_R+1)A_R^{-}+(n_R-1)A_R^{+}X_R^2},
\\
&&r_R'=-
\frac{(n_R+1)A_R^{+}+(n_R-1)A_R^{-}X_R^2}
{(n_R+1)A_R^{-}+(n_R-1)A_R^{+}X_R^2},\label{r_R'}
\end{eqnarray}\end{subequations}
where
\begin{equation}
A_R^{\sigma}=
\frac{1}{2}[(1+\sigma n_R)\sqrt{1-T}+ 1-\sigma n_R],
\quad X_R=e^{i\omega n_RL/2c}.
\end{equation}

We now apply the quantization condition (\ref{eq:rlrr}). This delivers the
equation
\begin{eqnarray}
&&(n_L+n_R)(A_L^{-}A_R^{-}-A_L^{+}A_R^{+}X_L^2X_R^2)
\nonumber \\
&&= (n_L-n_R)(A_L^{-}A_R^{+}X_R^2-A_L^{+}A_R^{-}X_L^2)
,\label{eq:qcond}
\end{eqnarray}
which has to be solved for the resonance frequencies entering via $X_L$ and
$X_R$.

Equation \eqref{eq:qcond} holds irrespective of whether the resonator is
$\mathcal{PT}$-symmetric or not, but from now on we assume that this symmetry
holds and therefore make use of $n_R=n_L^*=n_0-i n_I$ and
$A_R^\pm=(A_L^\pm)^*$. Moreover, we introduce the scaled dimensionless
frequency
\begin{equation}
\Omega=\omega n_0 L/c\label{eq:scaledomega}
\end{equation}
and the dimensionless degree of nonhermiticity
\begin{equation}
\alpha=n_I/n_0.\label{eq:alpha}
\end{equation}
In terms of these quantities, we can rewrite Eq.\ (\ref{eq:qcond}) as
\begin{equation}
|A_L^{+}|^2e^{i\Omega}-|A_L^{-}|^2e^{-i\Omega}
+ i\alpha(A_L^{-}{A_L^{+}}^*e^{\alpha\Omega}-A_L^{+}{A_L^{-}}^*e^{-\alpha\Omega})=0.
\label{eq:qcond2}
\end{equation}

\subsubsection{Closed system}

In the limit $T=0$ of a closed system, $A^\pm_{L,R}=1$. The quantization
condition (\ref{eq:qcond2}) then takes the form
\begin{equation}
\mathcal{F}(\Omega)\equiv \alpha \sinh(\alpha \Omega) +\sin(\Omega)=0,\label{rescond}
\end{equation}
which demands us to find the roots of a real function. The solutions are
real, or occur in complex conjugated pairs, as required by $\mathcal{PT}$
symmetry. We denote these solutions as $\bar\Omega_m$ (where
$m=1,2,3,\ldots$), so that we can refer back to them when we discuss the open
system.

The transition from a real to a complex spectrum is driven by the degree of
nonhermiticity $\alpha$. This transition involves exceptional points which
occur when two real frequencies coalesce, delivering the additional condition
\begin{equation}
\mathcal{F}'(\Omega)= \frac{\partial\mathcal{F}}{\partial\Omega} = \alpha^2 \cosh(\alpha \Omega) +\cos(\Omega)=0.\label{epcond}
\end{equation}
For an exceptional point, Eqs.~(\ref{rescond}) and (\ref{epcond}) have to be
fulfilled simultaneously. We denote the corresponding value of $\alpha$ and
the value $\bar \Omega$ of the two coalescing frequencies as
$\alpha_l^{\star}$ and $\bar\Omega_l^{\star}$, respectively, where
$l=1,2,3,\ldots$ enumerates the exceptional points. Equations~(\ref{rescond})
and (\ref{epcond}) then are equivalent to the conditions
\begin{equation}
\cos(\bar\Omega_l^{\star})=-\alpha_l^{\star},\quad
\cosh(\alpha_l^{\star}\bar\Omega_l^{\star})=1/\alpha_l^{\star}.
\label{epcondfinal}
\end{equation}

\begin{figure}[t]
\includegraphics[width=\columnwidth]{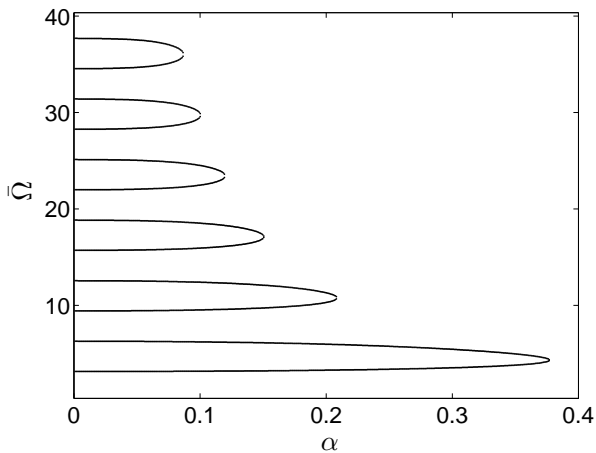}
\caption{Illustration of spontaneous $\mathcal{PT}$-symmetry breaking in the model
resonator of Fig.\ \ref{fig1}(c), for $T=0$ (closed system). The plot shows
the dependence of the 12 lowest-lying dimensionless resonance frequencies
$\bar\Omega$ on the hermiticity-breaking parameter $\alpha$, following their
trajectory as long as they are real. As $\alpha$ increases, adjacent
resonance frequencies approach each other pairwise, until they merge in an
exceptional point. Beyond the exceptional point, the involved frequencies
become a complex-conjugated pair, and then are no longer plotted. These
results are obtained by numerical solution of Eq.~\eqref{rescond}. }
\label{fig2}
\end{figure}

For illustration, we plot in Fig.~\ref{fig2} a set of quantized frequencies
as a function of $\alpha$, for a frequency range covering the 12 lowest-lying
levels. The figure displays six exceptional points at which pairs of real
frequencies coalesce. Beyond these points, the involved frequencies become
complex, and then are no longer plotted.

For $\alpha=0$, the $m$th resonance frequency is located at
$\bar\Omega_m=m\pi$ ($m=1,2,3,\ldots$). At the exception points two
consecutive frequencies $\bar\Omega_{2l-1}$ and $\bar\Omega_{2l}$ approach
the value $\bar\Omega_l^{\star}\approx 2l\pi-\pi/2$, an expression which
becomes more and more accurate as $l$ increases. In the same limit,
Eqs.~\eqref{rescond} and \eqref{epcond} deliver the approximate condition
\begin{equation}
\alpha_l^{\star} e^{\alpha_l^{\star} \bar\Omega_{l}^{\star}} \approx 2.
\label{eq:ep_large}
\end{equation}
With increasing $l$, $\alpha_l^{\star}$ decreases steadily, while the product
$\alpha_l^{\star} \bar\Omega_{l}^{\star}$ increases very slowly.

In order to investigate the behaviour close to an exceptional point, we
expand
\begin{equation}
{\cal F}(\alpha,\Omega)
\approx \frac{{\cal
F}''(\alpha_l^\star,\Omega_l^\star)}{2}(\Omega-\Omega_l^\star)^2+\dot{\cal
F}(\alpha_l^\star,\Omega_l^\star)(\alpha-\alpha_l^\star),
\label{eq:fexpand}
\end{equation}
where $\dot{\cal F}=\partial {\cal F}/\partial\alpha$, and we only kept the
leading nonvanishing terms in this expansion. Slightly below the $l$th
exceptional point, the two resonance frequencies are therefore given by
\begin{subequations}\begin{eqnarray}
\bar\Omega_{2l}(\alpha)&\approx&\bar\Omega_{l}^\star
+\sqrt{\frac{2\dot{\cal F}(\alpha_l^\star,\Omega_l^\star)}{{\cal F}''(\alpha_l^\star,\Omega_l^\star)}(\alpha_l^\star-\alpha)},\\
\bar\Omega_{2l-1}(\alpha)&\approx&\bar\Omega_{l}^\star
-\sqrt{\frac{2\dot{\cal F}(\alpha_l^\star,\Omega_l^\star)}{{\cal F}''(\alpha_l^\star,\Omega_l^\star)}(\alpha_l^\star-\alpha)}.
\end{eqnarray}\label{eq:omegaalpha}\end{subequations}
Because of the square-root dependence, the spectral arrangement near the
exceptional point occurs over a very small range of $\alpha$. In the
following, we will use the term ``far below (or above) the exceptional
point'' to indicate that the eigenvalues are well separated, while the regime
where they are close together is called ``slightly below (or above) the
exceptional point''. However, in terms of numerical values $\alpha$ actually
needs to approach $\alpha_l^{\star}$ very closely before the latter regime is
entered. For later convenience we also note that by a similar expansion as in
Eq.~\eqref{eq:fexpand}, the resonance splitting
\begin{eqnarray}
\bar\Omega_{2l}(\alpha)-\bar\Omega_{2l-1}(\alpha)&\approx&
2\mathcal{F}'(\alpha, \bar\Omega_{2l})/\mathcal{F}''(\alpha_{l}^{\star}, \bar\Omega_{l}^\star)
\label{eq:epdiff}
\end{eqnarray}
can be related to the deviation of $\mathcal{F}'$ from the condition
\eqref{epcond} for an exceptional point.

\begin{figure}[t]
\includegraphics[width=\columnwidth]{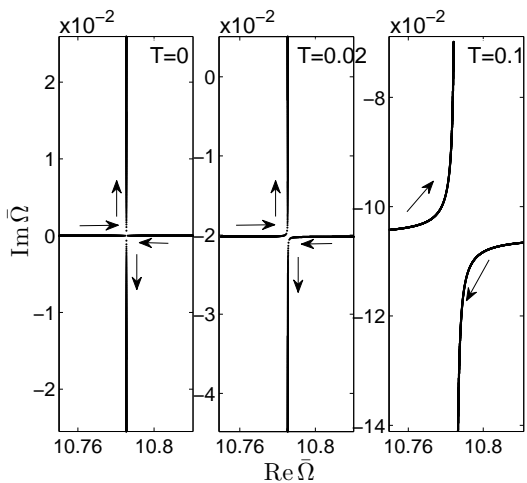}
\caption{Evolution of a pair of dimensionless resonance frequencies in the complex
plane, while sweeping the nonhermiticity parameter $\alpha$ of the
$\mathcal{PT}$-symmetric model resonator depicted in Fig.~\ref{fig1}(c).
Shown are the resonance frequencies of the third and fourth level, with focus
on the region around the exceptional point of the closed system ($T=0$, as
realized in panel a). Panels $(b)$ and $(c)$ show the lifting of exact
degeneracy when the system is slightly open ($T=0.02$ and $T=0.1$,
respectively). The arrows indicate the direction of the evolution with
increasing $\alpha$. Before the exceptional point, the resonances approach
each other roughly parallel to the real frequency axis. Beyond the
exceptional point, one of the two resonances is pushed towards the real axis,
which facilitates the reaching of the lasing threshold (see
Ref.~\cite{Y.D.Chong} for a system following the same scenario). These
results are obtained by numerical solution of Eq.~\eqref{eq:qcond2}, where we
set $n_0=2$.} \label{fig3}
\end{figure}

\subsubsection{Slightly open system}

When the resonator is opened, bound states turn into quasibound states with
complex resonance frequencies which are generally shifted by a mode-dependent
amount downwards in the complex plane. For the system studied here, we find
for $T\ll 1$ that this shift $\propto T$ is approximately rigid (i.e., mode
independent), and is combined with a lifting $\sim T^{3/2}$ of resonance
coalescence close to the exceptional points (to obtain exact coalescence in
the complex plane, additional parameters beside $\alpha$ would have to be
varied). These features follow by expanding the resonance condition
\eqref{eq:qcond2} up to order $T^2$, upon which it takes the form
\begin{equation}
\mathcal{F}(\Omega)+i\frac{\Gamma_0}{2} \mathcal{F}'(\Omega)
-\alpha(1 +\alpha^2)\frac{\Gamma_0^2}{8}\sinh\alpha\Omega =0
\label{approxres}
\end{equation}
where
\begin{equation}
\Gamma_0=\frac{n_0 T}{1-T/2}.
\end{equation}
To linear order in $\Gamma_0$, Eq.~\eqref{approxres} is solved by
\begin{equation}
\Omega_m=\bar\Omega_m-i\Gamma_0/2,\label{eq:omega}
\end{equation}
where $\bar\Omega_m$ are the bound-state frequencies of the closed system,
determined by Eq.~\eqref{rescond}. Slightly opening up this resonator thus
does not change the real parts of the resonance frequencies, and shifts the
imaginary parts by a mode-independent amount. In accordance to Eq.\
\eqref{eq:gamma3}, this shift does not depend on the degree of nonhermiticity
$\alpha$, which identifies $\Gamma_0$ as the cold-cavity decay rate. Notably,
$\Gamma_0$ also determines the rigid shift of complex-conjugated pairs if
$\mathcal{PT}$-symmetry in the closed system is spontaneously broken.

For illustration, we plot in Fig.~\ref{fig3} the evolution of two resonance
frequencies in the complex plane while the parameter $\alpha$ passes through
an exceptional point of the closed system. The arrows indicate the evolution
direction of the resonance frequencies with increasing $\alpha$. We set
$n_0=2$, such that the expected shift $\Gamma_0/2\approx T/(1-T/2)\approx T$
for small $T$. For the closed system (panel a), two real eigenvalues approach
each other horizontally until they merge at the exceptional point, after with
they become complex and move almost vertically away from each other. In
panels (b) and (c), where the leakiness is small ($T=0.02$) and moderate
($T=0.1$), respectively, the  resonance frequencies are shifted downwards by
the expected amounts (for $T=0.1$, $T/(1-T/2)=0.105$, which corresponds well
to the shift in the asymptotic regions to the left and right of the plotted
range). The exceptional point is lifted only very slightly in panel (b), but
much more distinctively in panel (c), and upon varying $T$ we find that the
distance of closest approach in the complex plane is indeed of order
$T^{3/2}$.

\begin{figure}[htb]
\includegraphics[width=.85\columnwidth]{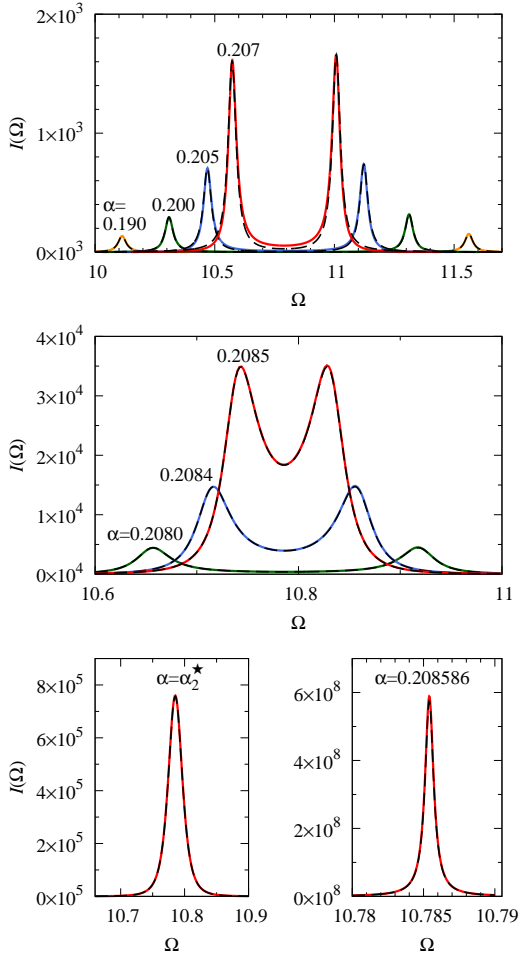}
\caption{(Color online) Frequency-resolved intensity
$I(\Omega)=I_L(\Omega)+I_R(\Omega)$ for the $\mathcal{PT}$-symmetric model
resonator depicted in Fig.~\ref{fig1}(c), for $n_0=2$ and $T=0.02$.
Frequencies are in the range of the third and fourth resonance. Exact results
from Eqs.~\eqref{I_L} and Eq.~\eqref{I_R} (solid curves) are compared with
various approximations (dashed curves), as the nonhermiticity parameter
$\alpha$ is changed to steer the system from the regime of well-isolated
resonances across the exceptional point  of the closed system
($\alpha_2^\star\approx0.208573$) up to just below the lasing threshold (the
threshold itself is at $\alpha\approx 0.208587$). Below the exceptional point
(top panel), the isolated resonances agree with the Lorentzian lineshape
\eqref{I_Lsimplepole}. Slightly below the exceptional point (middle panel),
the interfering resonances are well described by Eq.~\eqref{I_Ldoublepole}.
This expression also holds true for $\alpha=\alpha_2^\star$, where
Eq.~\eqref{I_Ldoublepole} degenerates into the squared-Lorentzian
\eqref{I_Lep} (lower left panel). Close to the lasing threshold one obtains
again a Lorentzian line shape, in agreement with \eqref{I_Lsimplepolebeyond}
(lower right panel). Note the dramatic increase in intensity as $\alpha$ is
changed incrementally.} \label{fig4}
\end{figure}

\subsection{Frequency-resolved output radiation intensity}\label{lineshape}

We now turn to the frequency-resolved output radiation intensities
$I_L(\omega)$ and $I_R(\omega)$ from the left and right openings of the
resonator. These follow by substituting the reflection and transmission
amplitudes \eqref{eq:sl} and \eqref{eq:sr}  into  Eqs.\ \eqref{I_L} and
\eqref{I_R}, respectively. Since these expressions contain the general
resonance quantization condition \eqref{eq:rlrr} in the denominator, the
resonance frequencies calculated in the previous section determine the
emission lines around which the output intensity is large. In this
subsection, we directly derive  expressions near resonance, including for the
situation near an exceptional point; in the following subsection these are
compared to the general near-resonant expression \eqref{finalresult}. For the
discussion, we again make use of the dimensionless variables $\Omega$ and
$\alpha$ [Eqs.~\eqref{eq:scaledomega} and \eqref{eq:alpha}, respectively].

Before we present analytical results, we illustrate the key features in
Fig.~\ref{fig4}, where the solid lines are numerical results for $I=I_L+I_R$,
obtained from the expressions \eqref{I_L} and \eqref{I_R} as described above
(the dashed lines represent analytical results derived below). We set
$n_0=2$, $T=0.02$, and plot the intensity in a frequency range covering the
third and the forth resonance, for values of $\alpha$ far below, near, and
slightly beyond the exceptional point $\alpha_2^{\star}\approx0.208573$ of
the closed system (reaching up very closely to the lasing threshold, which
occurs at $\alpha\approx 0.208587$).

The upper panel in Fig.~\ref{fig4} shows results for four values of $\alpha$
far below the exceptional point. The resonance frequencies are then well
isolated, and the output intensity displays well-resolved resonance peaks
which fit to a Lorentzian lineshape. As $\alpha$ increases, the resonances
draw together and increase in height, but do not change their width, which is
$\approx\Gamma_0$. The middle panel shows how resonances start to merge as
one approaches the exceptional point. This goes along with a dramatically
increasing intensity, which is in accordance to the expectation from
strongly-violated mode orthogonality. At the exceptional point  (lower left
panel), the resonances merge into a very high single peak, which can be
described by a \emph{squared} Lorentzian, which still is of width
$\sim\Gamma_0$. Moving beyond the exceptional point, the resonance peak
retains its centre. As one approaches the lasing threshold (where one of the
complex resonance frequencies approaches the real axis) the emission line
reverts to a Lorentzian, and the peak intensity increases further while the
resonance width decreases.

In order to explain these results we now obtain analytical expressions for
the resonance peaks in the output intensity, covering the whole range far below,
near and slightly beyond the
exceptional point (where one quickly reaches the lasing threshold). On
obtaining these expressions, it suffices to focus the attention on $I_L$, as
$I_R$ leads to the same results in the relevant leading orders in $T$. The
key step in the derivation is to approximate the resonant denominator in
Eq.~\eqref{I_L} by the same steps that lead from Eq.~\eqref{eq:qcond2} to the
simplified quantization condition \eqref{approxres}. Applying similar (more
straightforward) approximations also to the denominator, this leads to the
expression
\begin{equation}
I_L(\Omega)=\frac{1}{2\pi}\frac{\Gamma_0}{2}\frac{(1+\alpha^2)^2\sinh\alpha\Omega}
{|\mathcal{F}(\Omega)+i\frac{\Gamma_0}{2} \mathcal{F}'(\Omega)
-\alpha(1 +\alpha^2)\frac{\Gamma_0^2}{8}\sinh\alpha\Omega|^2},
\label{I_Lsimple}
\end{equation}
from which all subsequent results follow.

For an isolated resonance far below the exceptional point, we expand the
denominator around the resonance condition $\Omega=\bar\Omega_m$, where
$\bar\Omega_m$ is the real quantized frequency of the closed system, given by
$\mathcal{F}=0$ [Eq.~\eqref{rescond}]. This delivers the Lorentzian
expression
\begin{equation}
I_L(\Omega)=\frac{1}{2\pi}\frac{(1+\alpha^2)^2\sinh\alpha\bar\Omega_m}{\mathcal{F}^{\prime 2}(\bar\Omega_m)}
\frac{\Gamma_0/2}{(\Omega-\bar\Omega_m)^2+\Gamma_0^2/4}, \label{I_Lsimplepole}
\end{equation}
which in the upper panel of Fig.~\ref{fig4} is shown as a dashed line on top
of the solid lines representing the exact numerical results. In the hermitian
limit $\alpha\to 0$, $\mathcal{F}^{\prime}(\bar\Omega_m)\to 1$, and
\begin{equation}
I_L(\Omega)\approx I_L^{(0)}(\Omega)\equiv\frac{1}{2\pi}
\frac{\alpha\bar\Omega_m\Gamma_0/2}{(\Omega-\bar\Omega_m)^2+\Gamma_0^2/4}. \label{I_L0}
\end{equation}
As expected, Eq.~\eqref{I_Lsimplepole} diverges at an exceptional point,
which here is manifest because the term $\mathcal{F}'(\bar\Omega_m)$ then
vanishes [see Eq.~\eqref{epcond}]. This is remedied by keeping the next
orders in the expansion of the denominator in Eq.~\eqref{I_Lsimple}, which
results in the squared Lorentzian
\begin{eqnarray}
I_L(\Omega)&=&\frac{1}{2\pi}\frac{(1+\alpha_l^{{\star}2})^2\sinh\alpha_l^{\star}\bar\Omega_{l}^{\star}}
{\mathcal{F}^{\prime\prime 2}(\bar\Omega_{l}^{\star})}
\frac{2\Gamma_0}
{ |(\Omega-\bar\Omega_{l}^{\star})^2+\Gamma_0^2/4|^2}
\nonumber \\
&=&\frac{1}{2\pi}\frac{1}
{\alpha_l^{\star 2}\sinh\alpha_l^{\star}\bar\Omega_{l}^{\star}}
\frac{2\Gamma_0}
{ |(\Omega-\bar\Omega_{l}^{\star})^2+\Gamma_0^2/4|^2}
,
\label{I_Lep}
\end{eqnarray}
where we used Eq.~\eqref{rescond} to write
\begin{equation}
\mathcal{F}^{\prime\prime}(\Omega_{l}^{\star})=
\alpha_l^{\star}(1+\alpha_l^{{\star}2})\sinh\alpha_l^{\star}\bar\Omega_{l}^{\star}
.
\label{eq:fpp}
\end{equation}
Equation \eqref{I_Lep} is shown in the lower left panel of Fig.~\ref{fig4} as
the dashed curve on top of the numerical result at the exceptional point.

In order to better understand the approach to this squared Lorentzian, let us
rederive it by considering the merging of the two involved resonance
frequencies $\bar\Omega_{2l-1}$ and $\bar\Omega_{2l}$ of the closed system.
While each associated partial intensity diverges as one approaches the
exceptional point, their coherent sum
\begin{eqnarray}
&&\hspace*{-.5cm}I_L(\Omega)\approx\frac{1}{2\pi}
\frac{\Gamma_0}{2}\frac{(1+\alpha_l^{{\star}2})^2\sinh\alpha_l^{\star}\bar\Omega_{l}^{\star}}
{\mathcal{F}^{\prime 2}(\alpha,\bar\Omega_{2l})}
\nonumber \\ &&\qquad{}\times
\left|\frac{1}{\Omega-\bar\Omega_{2l-1}+i\Gamma_0/2}
-\frac{1}{\Omega-\bar\Omega_{2l}+i\Gamma_0/2}
\right|^2
\nonumber \\ &&
\hspace*{-.5cm}=\frac{1}{2\pi}\frac{1}
{\alpha_l^{\star 2}\sinh\alpha_l^{\star}\bar\Omega_{l}^{\star}}
\frac{2\Gamma_0}
{|(\Omega-\bar\Omega_{2l-1}+i\frac{\Gamma_0}{2})(\Omega-\bar\Omega_{2l}+i\frac{\Gamma_0}{2})|^2}
\nonumber \\
\label{I_Ldoublepole}
\end{eqnarray}
[where we have made use of Eqs.~\eqref{eq:epdiff} and \eqref{eq:fpp}] remains
finite, and reduces to the squared Lorentzian \eqref{I_Lep} when the two
resonance frequencies coalesce. Expression \eqref{I_Ldoublepole} is also
accurate slightly away from the exceptional point, as can be seen from the
curves in the middle panel of Fig.~\ref{fig4}.

Slightly beyond the exceptional point one approaches the lasing threshold,
where one of the two resonances (with index $2l-1$ if continuously labeled as
in Fig. \ref{fig3}) comes close to the real axis, thereby acquiring a much
reduced width $\Delta\Omega=-2\,{\rm Im}\,\Omega_{2l-1}\approx
\Gamma_0-2\,{\rm Im}\,\bar\Omega_{2l-1}\ll\Gamma$. The peak intensity of this
resonance then exceeds by far that of the other resonance, resulting again in
a (very narrow) Lorentzian
\begin{equation}
I_L(\Omega)= \frac{1}{2\pi}\frac{1}
{\alpha_l^{\star 2}\sinh\alpha_l^{\star}\bar\Omega_{l}^{\star}}
\frac{2/\Gamma_0}{(\Omega-{\rm Re}\,\bar\Omega_{2l-1})^2+\Delta\Omega^2/4}. \label{I_Lsimplepolebeyond}
\end{equation}
This is plotted in the bottom right panel of Fig.~\ref{fig4}.

\subsection{Implications of mode nonorthogonality}\label{sec:petermann}

We now discuss the preceding analytical results for the model resonator from
the perspective of mode nonorthogonality, based on the general considerations
presented in Sec.~\ref{sec:IV}. Our goal is to recover
Eq.~\eqref{I_Lsimplepole} by substituting the resonance wavefunction of the
model system into Eq.~\eqref{finalresult}.

Assuming no incoming radiation from the outside of the resonator, and
employing scaled units $s=x/L$, the wave function in the left and right parts
of the system (in which the respective refractive index is constant) can be
written as
\begin{subequations}
\begin{eqnarray}
&&\psi_L=r_{n_L,M}e^{i\Omega(1+i\alpha)(s+1/2)}+e^{-i\Omega(1+i\alpha)(s+1/2)}
,\\
&&\psi_R=a[e^{i\Omega(1-i\alpha)(s-1/2)}+r_{n_R,M}e^{-i\Omega(1-i\alpha)(s-1/2)}].\qquad
\end{eqnarray}
\end{subequations}
Here
\begin{subequations}
\begin{eqnarray}
&& r_{n_L,M}\approx -1+n_LT/2=-1+\frac{\Gamma_0}{2}(1+i\alpha)
,\\
&& r_{n_R,M}\approx -1+n_RT/2=-1+\frac{\Gamma_0}{2}(1-i\alpha),
\end{eqnarray}
\end{subequations}
are the reflection coefficients of the mirrors, including the step in the
refractive index from 1 to $n_L$ or $n_R$, respectively. The coefficient $a$
is determined by the matching condition $\psi_L(0)=\psi_R(0)$, giving for
$\Gamma_0\to 0$
\begin{equation}
a=\frac{\sinh[i\Omega(1+i\alpha)/2]}{\sinh[i\Omega(1-i\alpha)/2]}.
\end{equation}
The second matching condition $\psi_L'(0)=\psi_R'(0)$ recovers the
quantization condition \eqref{rescond}. From this, we only need the
previously established property $\Omega_m\approx\bar\Omega_m-i\Gamma_0/2$
[Eq.~\eqref{eq:omega}]. For $\Gamma_0\to 0$ and real $\Omega_m$, we
furthermore find $|a|=1$.

Based on these expressions, and keeping only leading orders in $\Gamma_0$
(which also appears in $\Omega_m=\bar\Omega_m-i\Gamma_0/2$), as well as using the quantization condition
\eqref{rescond},
we find
\begin{subequations}
\begin{eqnarray}
&&\int_{{\cal R}} {\rm Im}\,\Omega_m^2n^2 |\psi|^2
=-2 n_0^2(1+\alpha^2)\bar\Omega_m\sinh(\alpha\bar\Omega_m)
,\qquad
\\
&&\int_{{\cal R}+{\cal L}} {\rm Im}\,\Omega_m^2n^2|\psi|^2=
 -2n_0^2(1+\alpha^2)\bar \Omega_m\Gamma_0
,
\\
&&  \left|\int_{{\cal R}+{\cal L}} \Omega_m^2 n^2 \psi^2\right|^2=
[2\bar \Omega_m^2n_0^2{\cal F'}(\bar\Omega_m)]^2.
\end{eqnarray}
\end{subequations}

Equation~\eqref{finalresult} then delivers
\begin{equation}\label{finalresult2}
I(\Omega)=\frac{1}{2\pi}\frac{(1+\alpha^2)^2
\sinh(\alpha\bar\Omega_m)}{{\cal F'}^2(\bar \Omega_m)}
\frac{\Gamma_0}{(\Omega-\bar \Omega_m)^2+\Gamma_0^2/4},
\end{equation}
which indeed agrees exactly with our earlier result \eqref{I_Lsimplepole},
given that $I_L=I_R=I/2$ in the considered regime $\Gamma_0\ll 1$. It is noteworthy that we explicitly recovered  the term
$\mathcal{F}'$ in the denominator, which vanishes at an exceptional point.
This confirms that the integral in the denominator of \eqref{finalresult}
constitutes the appropriate overlap integral, which quantifies the degree of
mode nonorthogonality, as we already argued on general grounds in
Sec.~\ref{sec:IV}. Finally, we once more call attention to Eq.\
\eqref{I_Ldoublepole}, which shows how the intensity is regularized by the
interference of the near-degenerate resonances. In this construction, the
partial amplitudes are still consistent with Eq.~\eqref{finalresult2}. We
thus can conclude that the appearance of the squared-Lorentzian line shape at
an exceptional point can be explicitly linked to the self-orthogonality
property of the resonance wave function, Eq.~\eqref{selforth}.

\section{Conclusions}\label{sec:conclusion}
Motivated by the recent interest in optical nonhermitian
$\mathcal{PT}$-symmetric systems, we investigated the radiation intensity
emitted by a resonator which is partially filled with an amplifying and an
absorbing medium. This required to combine aspects of quantum noise with the
properties of the resonator modes (especially, the consequences of broken
hermiticity). We addressed these issues on the common basis of scattering
theory, which allows to include quantum noise via the quantum-optical
input-output formalism (see Secs.~\ref{sec:IIb} and \ref{sec:III}), while
also giving access to the resonant frequencies (see Sec.~\ref{sec:IIa}) and
partial amplitudes in different regions of the resonator (see
Sec.~\ref{sec:IV}).

Our main result is an expression of the near-resonant frequency-resolved
radiation intensity, Eq.~\eqref{finalresult}, which relates this quantity to
the properties of the resonant radiation mode, and includes an explicit
measure of mode nonorthogonality induced by the amplifying and absorbing
regions. $\mathcal{PT}$-symmetric systems provide natural access to
exceptional points, where two resonances become degenerate, and their modes
become self-orthogonal. The partial intensity of each resonance then
diverges, but their sum yields a finite result, with a squared-Lorentzian
line shape. Compared to the case of isolated resonances, we find that the
total intensity is dramatically increased (see Fig.~\ref{fig4}), which should
facilitate the observation of this radiation in experiments.

We validated these results for the case of a model resonator (see
Sec.~\ref{sec:V}), for which we obtained explicit analytical results in the
whole physically accessible range of broken hermiticity, from the case of
isolated resonances [Eq.~\eqref{I_Lsimplepole}] over the near-degenerate case
close to an exceptional point [Eq.~\eqref{I_Ldoublepole}] up to the lasing
threshold [Eq.~\eqref{I_Lsimplepolebeyond}].

\acknowledgments

GY and HSS gratefully acknowledge the support of NRF via grant 2009-0078437.


\begin{thebibliography}{98}
\bibitem{note1} This expression assumes that the amplification shifts the
    modes strictly upwards in the complex plane. For corrections due to the
    shift of the real part of the resonance frequency see Ref.~\cite{H.Schomerus1}.
\bibitem{note2} Note that for convenience we have inserted zero-width
    interfaces with refractive index $1$ between any two neighboring regions
    with different refractive index; this insertion does not alter the
    result.
\end{thebibliography}
\end{document}